\documentstyle[twocolumn,epsfig]{article}
\def\FT{Fault-\kern-2pt Tolerant}
\def\WWW{World-\kern-1pt Wide Web}

\makeatletter
\def\@normalsize{\@setsize\normalsize{10pt}\xpt\@xpt
\abovedisplayskip 10pt plus2pt minus5pt\belowdisplayskip \abovedisplayskip 
\abovedisplayshortskip \z@ plus3pt\belowdisplayshortskip 6pt plus3pt
minus3pt\let\@listi\@listI}
\def\subsize{\@setsize\subsize{12pt}\xipt\@xipt}
\def\section{\@startsection {section}{1}{\z@}{1.0ex plus 1ex minus
 .2ex}{.2ex plus .2ex}{\large\bf}}
\def\subsection{\@startsection {subsection}{2}{\z@}{.2ex plus 1ex} 
{.2ex plus .2ex}{\subsize\bf}}
\makeatother
\begin{document}
\date{}
\title{\Large\bf A Hypermedia Distributed Application\\
for Monitoring and Fault-Injection in\\
Embedded \FT{} Parallel Programs}
\author{V. De Florio\ \ \ \ G. Deconinck\ \ \ \ M. Truyens\ \ \ \ 
W. Rosseel\ \ \ \ R. Lauwereins \\ \\
Katholieke Universiteit Leuven\\
Electrical Engineering Dept. -- ACCA\\
Kard. Mercierlaan 94 -- B-3001 Heverlee -- Belgium}
\maketitle

\thispagestyle{empty}
\subsection*{\centering Abstract}
{\em
We describe a distributed, multimedia application
which is being developed in the framework of the
ESPRIT-IV Project 21012 EFTOS (Embedded \FT{} Supercomputing).
The application dynamically sets up
a hierarchy of HTML pages reflecting the current status of an
EFTOS-compliant dependable application running on a Parsytec CC system.
These pages are fed to a \WWW{} browser playing
the role of a hypermedia monitor.  The adopted approach allows the user
to concentrate on the high-level aspects of his/her application so to
quickly assess the quality of its current fault-tolerance design. This view
of the system lends itself well for being coupled with a tool
to interactively inject software faults in the user application; 
this tool is currently under development.
}
\section{Introduction}
As systems get more and more complex, the need for a one-look snapshot of
their activity is indeed ever increasing. This need has been strongly
felt by people involved in the development of the ESPRIT-IV Project
EFTOS~\cite{DDLV97} (Embedded \FT{} Supercomputing), whose
aim is to set up a software framework for integrating fault-tolerance into
embedded distributed high-performance applications in a flexible and easy way.

Through the adoption of the EFTOS framework, the target
application running on a parallel computer is plugged into a hierarchical,
layered system whose structure and basic components are:

\begin{itemize}
\item at the lowest level, a set of parametrisable functions managing error
detection (Dtools)  and error recovery (Rtools). A typical Dtool is a
watchdog timer thread or a trap-handling mechanism;  a Rtool is e.g., a
fast-reboot thread capable of restarting a single node or a set of nodes.
These are the basic components that are plugged into the embedded
application to make it more dependable. EFTOS supplies a number of these
Dtools and Rtools, plus an API for incorporating user-defined EFTOS-compliant
tools; 
\item at the middle level, a distributed application called {\em DIR net\/}
(detection, isolation, and recovery network) is available to coherently
combine Dtools and Rtools, to ensure consistent strategies throughout the
whole system, and to play the role of a backbone handling information 
to and from
the fault-tolerance elements. To fulfill these requirements, the DIR net
makes use of processes called Manager, Agents, and Backup Agents;
\item at the highest level, these elements are combined into dependable
mechanisms e.g., methods to guarantee fault-tolerant communication, voting
methods and so on.  
\end{itemize}

During the lifetime of the application, this framework guards the
application from a series of possible deviations from the expected
activity; this is done by executing detection, isolation, and reconfiguration
tasks. For instance, a protection violation caught in a thread by a
trap handling Dtool may trigger a relocation of that thread elsewhere
in the system. As another example, if an error is detected which affects
a component of the DIR net itself, say a Manager, then the system will
isolate that component and elect another one (actually, one
of the Backup Agents) as the DIR Manager.

To let the user keep track of events like those sketched above, 
the DIR net continuously prints on the system console short textual descriptions.
Evidently such a linear, unstructured listing of events pertaining
different aspects of different actions taking place in different
points of the user application, do not make up the best mechanism to
gain insight in the overall state of the fault-tolerant system.
On the contrary, a hierarchical, dynamic view of the structure
and behavior of this system, including:

\begin{itemize} 
\item its current shape (on which node which components are
running, and their topology), 
\item the current state of its components
(for instance, whether they are regarded to be correct, faulty,
or are being recovered), 
\item each component's running history,
\end{itemize} 

appeared to be the best solution fulfilling our needs.

Two main advantages from the adoption of such a strategy were foreseen, namely:
\begin{itemize} 
\item (at design and system validation time) the possibility
to assist the user assessing
and/or validating his/her EFTOS-based fault-tolerance design, 
\item (at run time) the possibility
to shorten the latency between the occurrence of the event, its
comprehension, and a proper reaction at user level\footnote{%
Indeed, the high volume of data coming out of such a complex system is very
likely to at least delay the appearance of the failure in the so-called
{\em user's universe\/} i.e., ``where the user of a system ultimately sees
the effect of faults and errors'' \cite{John89a}; in some cases it may also
make it transparent to the user altogether.}.
\end{itemize}

This work describes the architectural solution that has been successfully
adopted within EFTOS to easily and quickly develop a tool fulfilling
the above stated needs---a portable, highly
customizable hypermedia monitor for the EFTOS applications making use of
cheap, ready available off-the-shelf software components like e.g., the Netscape Navigator.  
It also shows how this monitor supplies the user with the needed structured
information, and how it proved its usefulness within EFTOS.  In particular
an extension is described, currently under development, by means of
which our monitor is turned into a versatile tool for 
fault-injection. 
\typeout{A list of future developments conclude this article??}

\section{Design Requirements}\label{req}

In order to quickly deliver human-comprehensible information from the
gigantic data stream produced by an EFTOS application, two needs have
been assessed:

\begin{itemize} 
\item a hierarchical representation of the data.

(Most of the produced data is available, but it has to be
organized and made browsable in ``layers'':

\begin{itemize}
\item at the highest level, only the logical structure of the application
should be displayed: which nodes are used, the EFTOS components executed on
each node, and their overall state;
\item at a medium level, a concise description of the events pertaining each particular
node should be made available;
\item at the lowest level, a deeper description of each particular event may also be
supplied on user-demand), 
\end{itemize}

\item the use of multimedia.

``An image is worth a thousand words'', they say, and maybe even more
insight can be derived from the extensive use of colors, sounds,
video-clips and so on. For instance, re-coloring a green image to red
may lead the user into realizing that a previously good situation has
turned into a problem. The use of colors traditionally associated to
meanings, or whose meaning can be borrowed from well-known everyday
situations (e.g., those of traffic-lights) further speeds up the
delivery of the information to the user. 
\end{itemize}

Both things are available nowadays in products like Netscape or similar
\WWW{}~\cite{www} browsers which are able to render hierarchies of
dynamically produced HTML~\cite{html} pages. We therefore decided to
develop a distributed application piloting a WWW browser which in turn
plays the role of a hypermedia renderer for the EFTOS system activity. 
This product,
which we call the EFTOS Monitor, is described in the following Sections.

\section{The Architecture of the EFTOS Monitor}
\begin{figure}[t]
\psfig{file=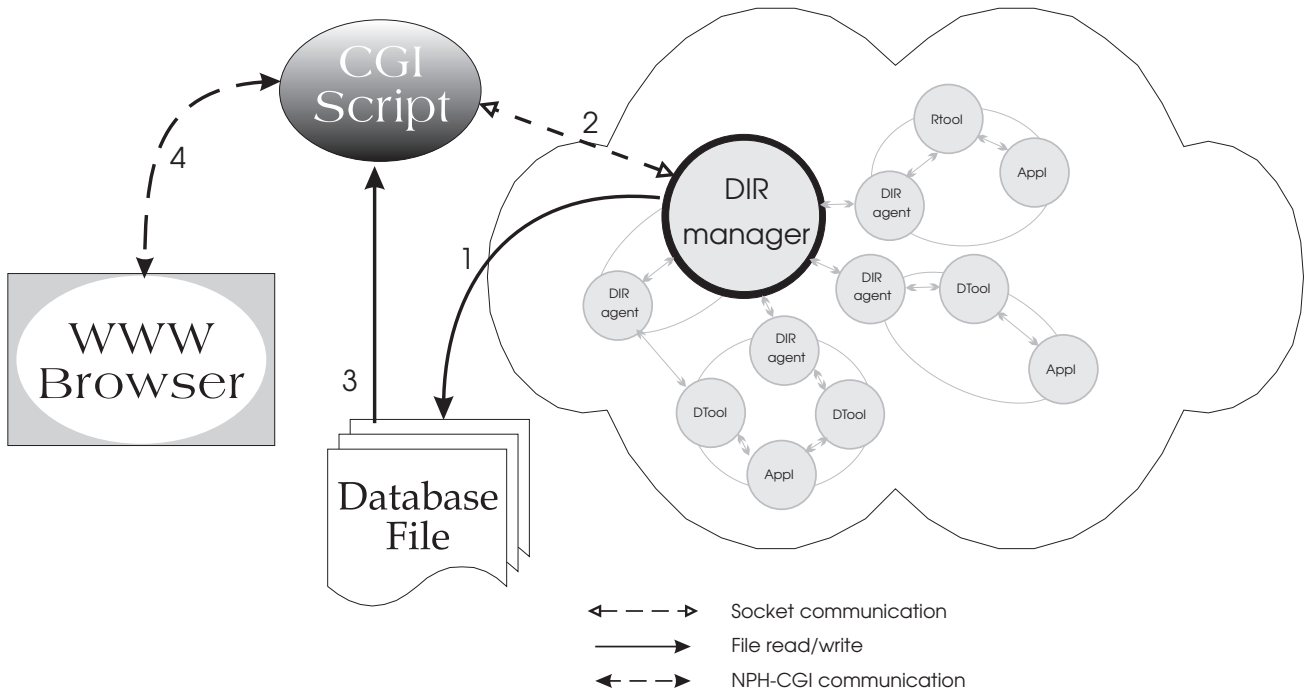,width=6.8in}
\caption{\sf The architecture of the EFTOS Monitor: the CGI scripts and the EFTOS
application share the same file system and communicate via a socket
stream: each time a new event takes place, the DIR net updates a special
database (1) and sends a notification to the main CGI script (2).
The latter reads the database (3) on
the arrival of the  notification and converts it into a
HTML hypertext, which is then fed (4) to a Netscape or another \WWW{}
browser for hypermedia rendering. 
The client part of the Monitor is integrated in the DIR net
Manager process.}\label{mon.arch}
\end{figure}

The EFTOS Monitor basically consists of three components 
(see Fig.~\ref{mon.arch}):

\begin{enumerate}
\item a {\em client\/} module, to be run by the DIR net;
\item an ``{\em intermediate\/}'' module, to be 
run by a number of Common Gateway Interface~\cite{cgi} (CGI) scripts;
\item the ``{\em renderer\/}'' i.e., a \WWW{} browser.
\end{enumerate}

\begin{itemize} 
\item The client part, together with the DIR net and the user application,
runs on a Parsytec CC system~\cite{Pars96a}, a distributed-memory MIMD
supercomputer consisting of powerful processing nodes 
based on PowerPC 604
at 133MHz, dedicated high-speed links, I/O modules and routers. 
The system adopts the thread processing model; threads exchange messages
through a proprietary message passing library called EPX~\cite{Pars96b}
({\em Embedded Parallel extensions to uniX\/}).
The main tasks of the client module are the set up and the management 
of a database maintaining an up-to-date snapshot of the
system activity, including the current mapping of the DIR net's components
onto the processing nodes and the state and current activity of each
component. This module also connects to the intermediate part 
via TCP sockets (see for instance~\cite{comer3}) and 
signals it on the 
very beginning and on the occurrence of each state
\linebreak

\vspace*{8.854cm}
transition.
\item The intermediate module consists of a hierarchy 
of CGI scripts spawned by an Apache HTTP~\cite{http10,http11} 
daemon, all running on the workstation hosting
the CC system.  The root script of this hierarchy
connects with the client module and acts as a TCP server: for each new
stimulus, the snapshot file is read over and a HTML document is produced
and fed to the renderer.  A connection is also started up with this latter
so to be able to tightly interact with it without the intervention of the
HTTP daemon: having done like this, one CGI script may stay alive and
produce multiple HTML requests, which is not the case in ordinary
CGI script---this special feature is known as
``non-parsing header'' (NPH) mode~\cite{cgi}.  Logically speaking, we may
say that the intermediate module acts as a gateway between the CC system
and the hypermedia renderer.  Like mythical Janus (It. {\em Giano\/}), 
one face is turned to
the client module and gathers its requests, while the other is turned to the
renderer and translates those requests in HTML---its main component has
therefore been called {\tt cgiano}.

\item The third component, the renderer, simply is a browser like Netscape
playing the role of a server able to display HTML data.
\end{itemize}

The application is started in two steps via a shell script whose first
task is to run the renderer (or to reconnect to a previously run renderer:
this latter is possible using e.g., the remote control
extensions~\cite{rc} of Netscape, or an approach based on the Common Client
Interface~\cite{dev29} mechanism of Mosaic; see for instance~\cite{mudhoney}).
The renderer is run with a uniform
resource locator~\cite{url} (URL) pointing to the root-level CGI script,
which connects to Netscape in NPH-mode and starts listening for a TCP
connection. 

\begin{figure}
\psfig{file=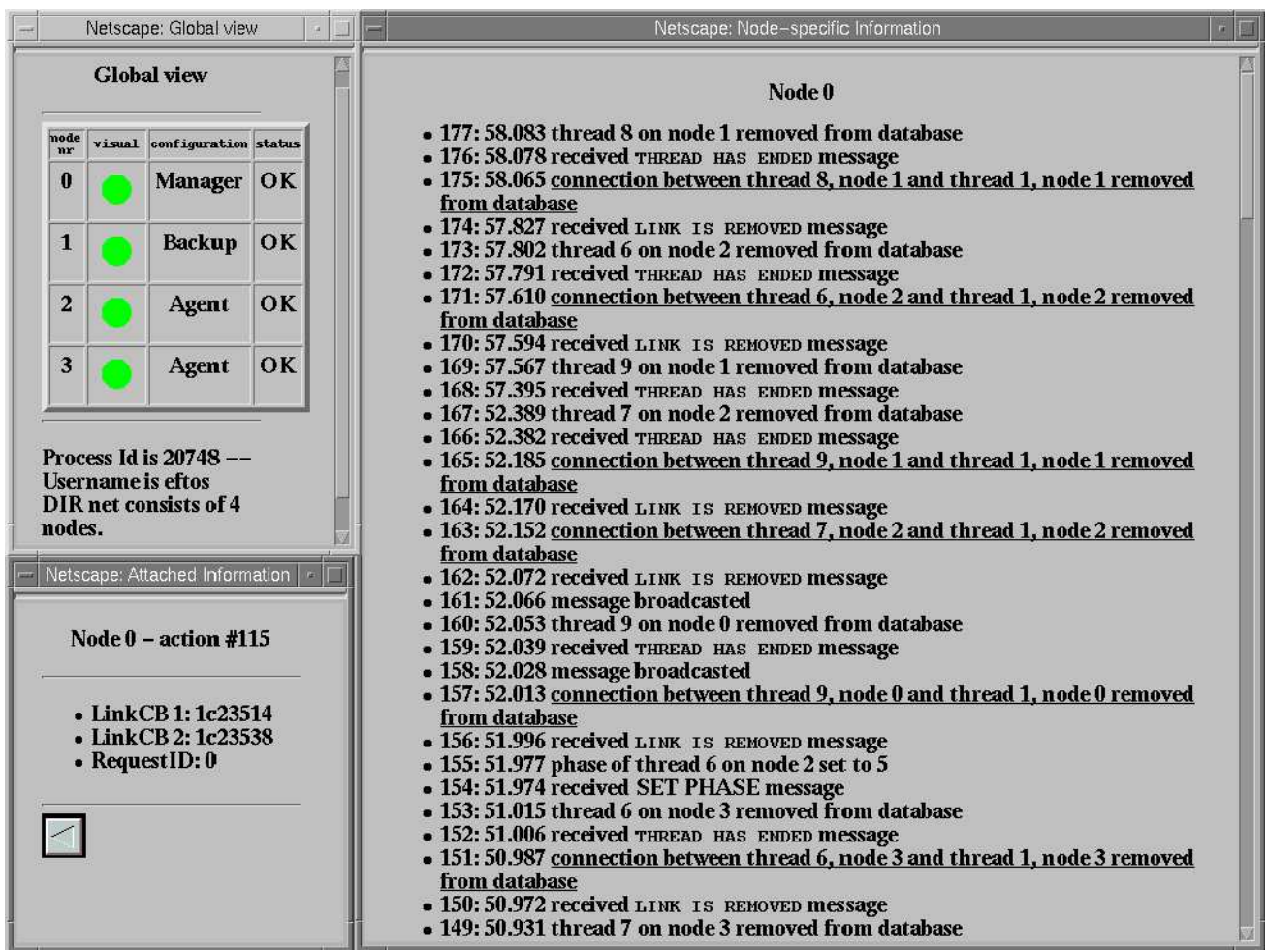,width=6.8in}
\caption{\sf The three windows of the EFTOS Monitor. In window 1
(``Global view''), the {\tt visual} column
contains graphical hyperlinks pointing to second-level information about
the corresponding processing node at the same row. {\tt Configuration} is
the DIR net-role.  {\tt Status} may be one of
the following values: {\tt OK}, {\tt Faulty}, {\tt Isolated},
{\tt Recovering}, and {\tt Killed}.
Some minor information is also displayed at
the bottom of the page.  The right-hand hypertext (window ``Node-specific
information'') is the result of
``clicking'' on the top circular icon and enumerates the actions that have just
taken place on node 0, fresher-to-older. The elapsed time (in seconds)
corresponding to each event is
displayed. Underlined sentences may be further expanded by clicking on
them e.g., the bottom-left image reports about action number 115 of the
hypertext.}\label{three.ps} 
\end{figure}

As a second step, the shell script spawns the parallel application on the
CC system.  Then the application launches the DIR net and the Monitor's
client module; this latter initializes the snapshot files, connects to the
CGI script and sends it the first signal. The script reacts to that
stimulus by translating the main snapshot file in HTML and requesting the
renderer to display it.  The top-left image in Fig.~\ref{three.ps}
shows a typical output of this phase:  the EFTOS application appears to
the user as an HTML table depicting the processing nodes in the user
partition. 
The state of each module
is illustrated by means of colors with obvious meaning (green is
``OK'', red is ``not OK'', yellow
means that the module is currently being
recovered, and so on). In this way the user can immediately perceive
whether a node is ready or not and which actions are carried out on it, as
asked for in the requirements (\S\ref{req}.)

Information displayed in this HTML document only covers the logical
structure of the application.  If the user ``clicks'' any icon on this
page, a high-level hypertextual description of the DIR net-events
pertaining that specific node is displayed in a separate, cloned
Netscape session (see Fig.~\ref{three.ps}, Window ``Node-specific
Information''.) To keep
this page up to date, an automatic reload is periodically performed.
This technique is explained e.g., in~\cite{cgi}.

This secondary document is in turn a hypertext whose links point to
in-depth descriptions of each specific event (see Fig.~\ref{three.ps},
Window ``Attached Information.'')

\section{Architecture Assessment}
A number of observations may be drawn upon the above presented architecture;
in particular:
\begin{itemize}
\item in our experience the architecture is easy and fast to design and develop, and effective 
especially towards fast prototyping;
\item it is based on unmodified, low-cost, off-the-shelf hypermedia components
which are widely available, continuously supported and updated on a wide range of
hardware architectures;
\item it is open, in the sense that the architecture is based on wide-spread standards
e.g., the use of uniform resource locators~\cite{url} within a World-Wide Web 
interconnection, the HTML language, TCP/IP sockets, the MIME classification, and so on;
\item it is distributed, and in particular the renderer may run on any X11-compliant
Display Server, including a remote PC.
\end{itemize}

A possible alternative is to develop a custom application to play as a tailored monitoring
tool. As an example, Scientific Computing Associates' TupleScope visual debugger is a
custom X-based visualisation and debugging tool for parallel programs using 
the LINDA approach~\cite{CaGe1}. This may result in higher performance and
possibly be more flexible but of course:

\begin{itemize}
\item it reasonably requires more time to develop even for a simple prototype;
\item it requires custom design and development choices that may impact portability
and supported features e.g., which software development environment and specifically
which language and which libraries to use, or whether to restrict the hypermedia
rendering to images or to use sounds as well---these choices may be simply skipped
in our approach;
\item distribution and hypermedia issues call for specific support which turn into
higher costs and longer times.
\end{itemize}

For instance, TupleScope runs with the user application by adding a special
linking option at compile time to the user application; this means it has been
developed on purpose as a custom X11 application.
Though it perfectly addresses its own goals, it has limited rendering capabilities 
(it only deals with static images) and would certainly require non-negligible
efforts to adapt it towards other media. Moreover, TupleScope is available on a
number of platform, though the costs of this portability and consequent support
are certainly not negligible as well.

\section{A Tool for Fault Injection}
The same approach used to monitor the state of an EFTOS-compliant
application is also effective in order to actively interact with it. Considering once again
\linebreak

\vspace*{14.1cm}

\noindent
Fig.~\ref{mon.arch}, a control path may
be drawn starting 
from the user at his/her browser, then crossing a CGI
script, and eventually reaching the user application. 
It is therefore
fairly possible to add a layer to the hierarchy of HTML pages dynamically
created by the intermediate modules so that the user may freely choose
among a certain set of malicious actions to bring against an EFTOS
application, including for instance: 
\begin{itemize}
\item an integer division-by-zero,
\item a segmentation violation,
\item a link failure,
\item rebooting a processing node,
\item killing a thread.
\end{itemize}
These requests would then reach a CGI script, be translated in appropriate
system- or application-level actions, which would then be executed or turned into
fault-injection requests to be fulfilled by the DIR Manager.
As an example of system-level action, the CGI script may directly execute
a system command to reboot one node in the CC system.
As of application-level actions, the Manager may
for instance ask the trap handler tool to trigger
a specific signal like {\tt SIGSEGV} (segmentation violation) on a certain
thread; or it may request a watchdog timer tool on a particular node to behave
like if it had detected a time-out.

\begin{figure}
\centerline{\psfig{file=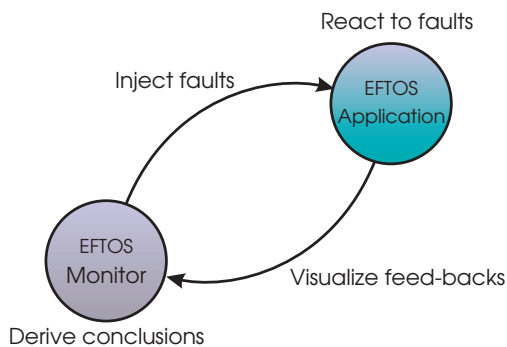,width=2.6in}}
\caption{\sf The recursive loop of fault injection and monitoring.}
\label{loop.eps}
\end{figure}

As a direct consequence of the injection of these faults, a number of
detection, isolation, and recovery actions will take place on the system
according to the EFTOS-based fault-tolerance strategies adopted by the
designer in his/her application. These actions will then be reported in
the snapshot files and displayed by the Monitor. 
This process, summarized in Fig.~\ref{loop.eps}, may be modeled after a
recursive loop like follows:

\noindent\hrulefill
\begin{quote}
  {\sf do \{} \\
    \hspace*{12pt}  Inject fault; \\
    \hspace*{12pt}  Observe feed-back;\\
    \hspace*{12pt}  Derive conclusions;\\
    \hspace*{12pt}  Correct the fault tolerance model;\\
  {\sf \} while} (model is unsatisfying).
\end{quote}
\noindent\hrulefill

In our opinion this procedure should result in an
extremely useful tool for rapidly assessing a design, trying alternative
fault-tolerance strategies, and overloading the system with malicious attacks
aiming at verifying its resilience, with a quick and meaningful feed-back
from the system. 

\section{Conclusions}
We presented the current state of development of a distributed 
application for monitoring the fault tolerance aspects of an embedded
parallel application and for interactively injecting faults into it.
The overall system makes up an integrated environment in which
they cyclically evolve: the application, a sophisticated
graphical rendering of the results, and real-time interactions
such that the researcher is made able to verify the hypothesis he/she
is formulating about the system. 

The design choice to adopt low-cost, off-the-shelf components for hypermedia
rendering revealed to be cost-effective, to speed up the development
process, to match the design requirements, and to point at
more ambitious capabilities and features. In particular, the use of
a \WWW{} browser as hypermedia renderer paves the way for 
future client-based extensions based on JavaScript or
Java~\cite{java}, and lets our application inherit the benefits
of the volcanic evolutions of the HTML languages, the HTTP protocol,
multimedia capabilities of the browsers, and so on. 

The high degree of openness proven by this heterogeneous application
basing itself on uniform communication mechanisms and standardized
access interfaces guarantees portability and makes it also a good
starting point towards
the development of similarly structured applications ranging from
remote equipment control to hypermedia multi-user environments.

We are currently using our Monitor during the development of
the new versions of the EFTOS fault tolerance framework.
The deeper insight that we have gained from it on the run-time aspects 
of our applications has turned into an invaluable tool to speed up
our development phases.

\subsection*{Acknowledgements}
This project is partly sponsored 
by an FWO Krediet aan Navorsers, by the Esprit-IV
Project 21012 EFTOS, and by COF/96/11. 
Vincenzo De Florio is on leave from Tecnopolis CSATA Novus Ortus.  
Geert Deconinck is a Postdoctoral Fellow of the Fund for Scientific Research 
- Flanders (Belgium) (F.W.O.).
Rudy Lauwereins is a Senior Research Associate of F.W.O.

\end{document}